%% file: template.tex
\documentclass[a4paper]{article}

\usepackage{INTERSPEECH2021}

\usepackage{subfig} %
\usepackage{float} %
\usepackage{booktabs} %

\title{Target Confusion in End-to-end Speaker Extraction: Analysis and Approaches}
\name{Zifeng Zhao$^1$, Dongchao Yang$^1$, Rongzhi Gu$^1$, Haoran Zhang$^1$, Yuexian Zou$^{1,*}$\thanks{* Corresponding author.}\thanks{This paper is supported by the Shenzhen Science \& Technology Fundamental Research Programs (No:JSGG20191129105421211 and GXWD20201231165807007-20200814115301001).}}

\address{$^1$ADSPLAB, School of ECE, Peking University, Shenzhen, China}
\email{\{zhaozifeng, dongchao98\}@stu.pku.edu.cn, \{1701111335, haoranzhang, zouyx\}@pku.edu.cn}

\begin{document}

\maketitle
\begin{abstract}
    Recently, end-to-end speaker extraction has attracted increasing attention and shown promising results. However, its performance is often inferior to that of a blind source separation (BSS) counterpart with a similar network architecture, due to the auxiliary speaker encoder may sometimes generate ambiguous speaker embeddings. Such ambiguous guidance information may confuse the separation network and hence lead to wrong extraction results, which deteriorates the overall performance. We refer to this as the \textit{target confusion problem}. In this paper, we conduct an analysis of such an issue and solve it in two stages. In the training phase, we propose to integrate metric learning methods to improve the distinguishability of embeddings produced by the speaker encoder. While for inference, a novel post-filtering strategy is designed to revise the wrong results. Specifically, we first identify these confusion samples by measuring the similarities between output estimates and enrollment utterances, after which the true target sources are recovered by a subtraction operation. Experiments show that performance improvement of more than 1dB SI-SDRi can be brought, which validates the effectiveness of our methods and emphasizes the impact of the \textit{target confusion problem}\footnote{A demo is available at https://zhazhafon.github.io/demo-confusion/}. 
    
\end{abstract}

\section{Introduction}

    Speech separation, also referred to as the cocktail-party problem, is considered to be one of the fundamental problems in speech processing areas\cite{Review}. Although easy for human beings, the same task is still challenging for machines.
    
     A speaker extraction model based on deep neural network (DNN) consists of two parts: a speaker encoder, which maps the enrollment utterance of the target speaker to an embedding, and a separation network, which extracts the target speaker's speech from the mixture under the guidance of the injected speaker embedding. In particular, these two components are jointly trained from scratch in end-to-end speaker extraction. Many studies developed their deep models based on state-of-the-art separation network architectures from BSS (e.g. TCN\cite{SpEx}\cite{TD-SpeakerBeam}\cite{DPRNN-SpkExtr} and DPRNN\cite{DPRNN-SpkExtr}\cite{NewInsights}), and achieved considerable performance.
    
\begin{figure}[t]
      \centering
      \includegraphics[width=\linewidth]{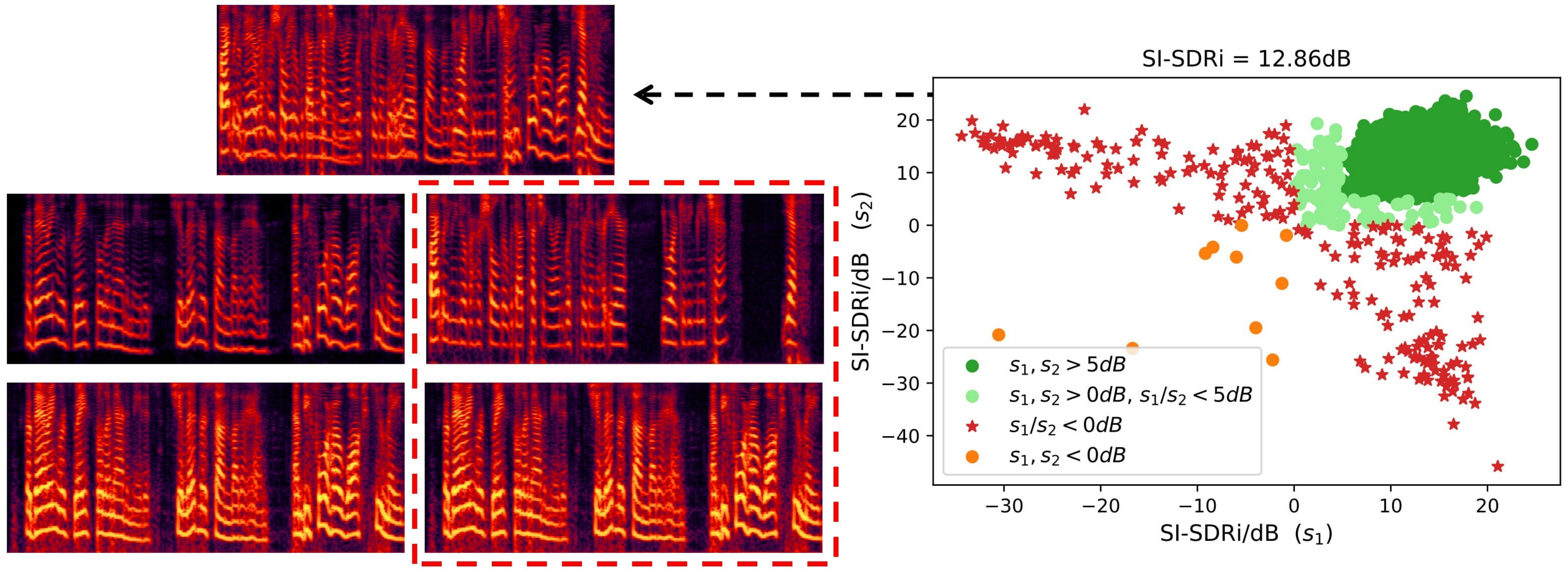}
      \caption{The \textit{target confusion problem} in a two-speaker scenario. The scatter diagram on the right is the performance distribution in terms of SI-SDRi on the test set of Libri2Mix, each axis corresponds to a speaker. $s_1, s_2>5dB$ denotes that SI-SDRi of both $s_1$ and $s_2$ are above 5dB, while $s_1/s_2<5dB$ means that either of them is below 5dB. Spectrograms on the left are from a data sample where target confusion happens. On the top is the observed mixture. Spectrograms in the second row are the ground truths for each target speaker, and those in the last row are the output estimates of each target source.}
      \label{fig:LongTail}
    \vspace{-2em} 
    \end{figure}

    However, our preliminary experiments as well as recent research\cite{NewInsights} show that, the performances of end-to-end speaker extraction are prone to long-tail distributions, which is depicted in Figure \ref{fig:LongTail}. As a result, end-to-end speaker extraction is often slightly inferior to its BSS counterpart when a similar separation network is adopted\cite{TD-SpeakerBeam}\cite{NewInsights}, despite the assistance of an additional speaker encoder and enrollment utterances. Such a gap originates from an issue which we term as the \textit{target confusion problem} in this paper, where the speaker embedding provides an ambiguous guidance, and thus the separation network targets at a wrong speaker (i.e. the interferer). This is illustrated in the red dashed box in Figure \ref{fig:LongTail}. Intuitively, there are two possible causes for this phenomenon. One is the \textit{utterance bias}, that is, the target speaker's speech (either the source or the enrollment) in the data sample deviates from its speaker cluster; The other is the \textit{embedding bias}, which means the output of speaker encoder does not  represent the guidance information accurately.
    
    In previous studies, multi-class cross-entropy (CE loss) was proposed specially for the speaker encoder in end-to-end speaker extraction, and joint training is carried out together with the reconstruction loss through a multi-task learning\cite{SpEx}\cite{TD-SpeakerBeam}\cite{SpEx+}\cite{SpEx++}. However, such a classification paradigm does not optimize similarities explicitly, which may produce suboptimal embeddings for end-to-end speaker extraction.
    
    \begin{figure*}[t]
      \centering
      \subfloat[Measured by ECAPA-TDNN]{
        \label{fig:ECAPA-a}
        \includegraphics[height=2.2cm]{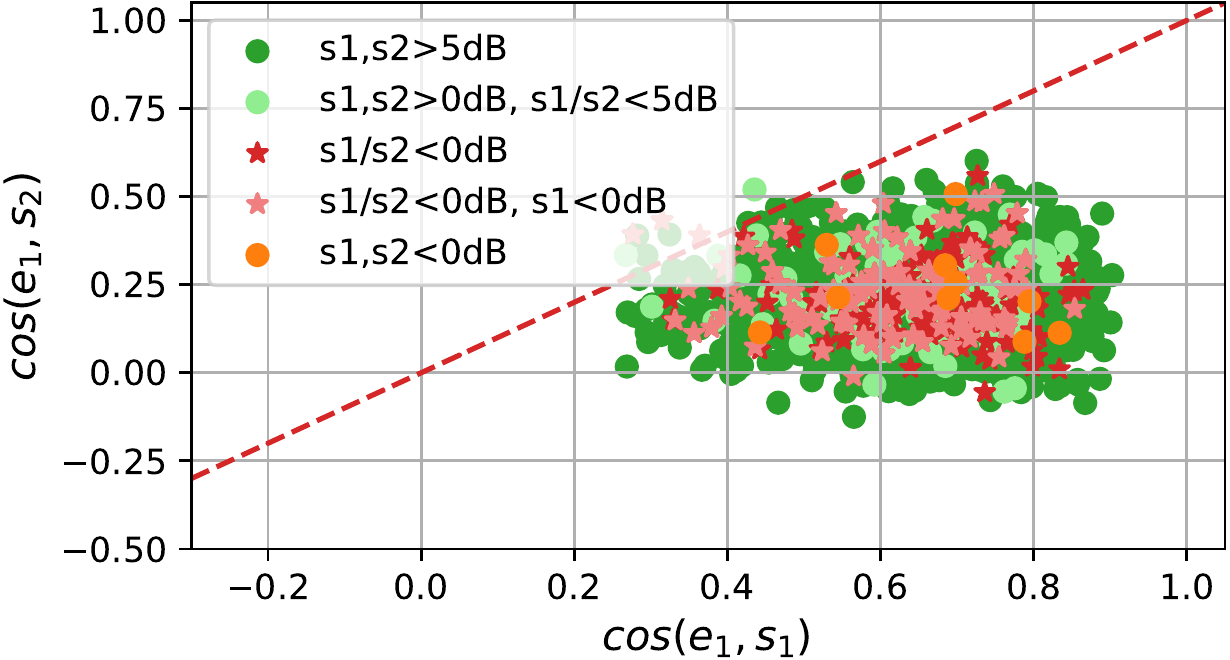}}
        \hspace{0.2em}
      \subfloat[Measured by ECAPA-TDNN]{
        \label{fig:ECAPA-b}
        \includegraphics[height=2.2cm]{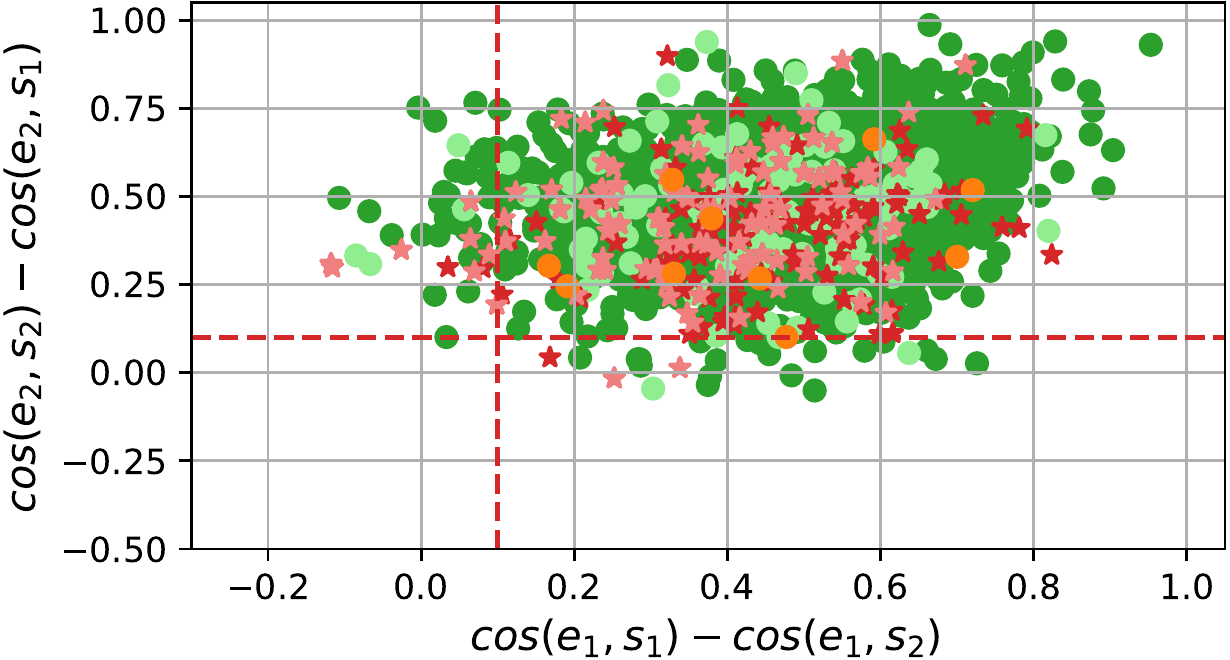}}
        \hspace{0.2em}
      \subfloat[Measured by speaker encoder]{
        \label{fig:SpeakerEncoder-c}
        \includegraphics[height=2.2cm]{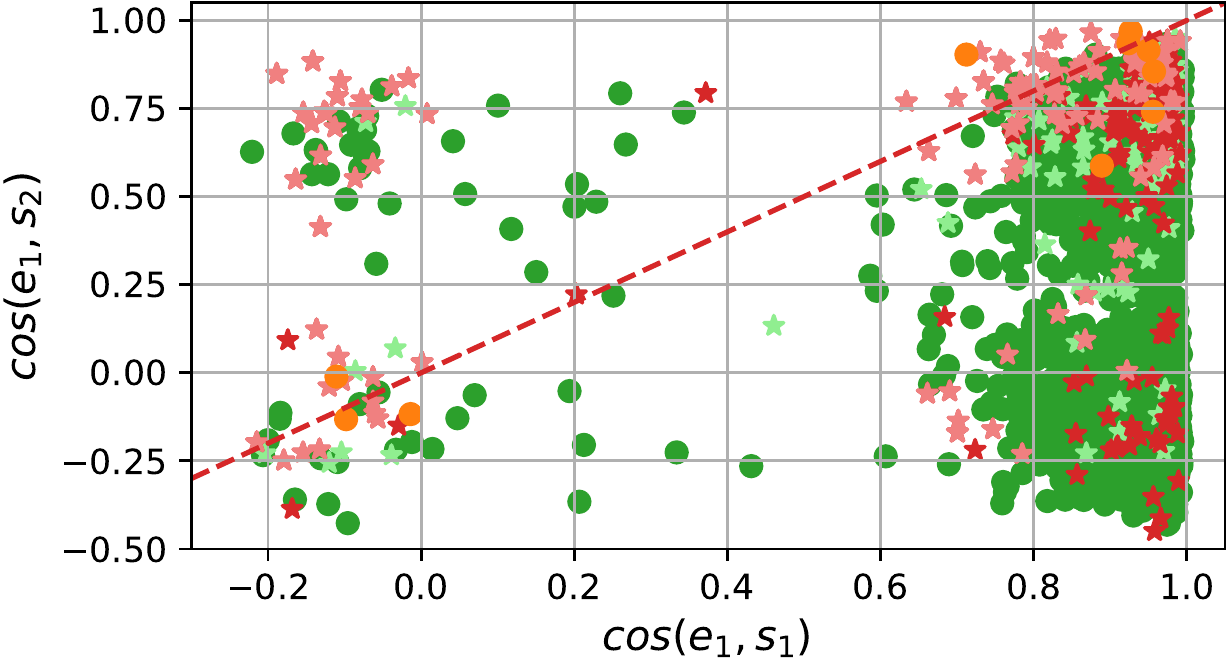}}
        \hspace{0.2em}
      \subfloat[Measured by speaker encoder]{
        \label{fig:SpeakerEncoder-d}
        \includegraphics[height=2.2cm]{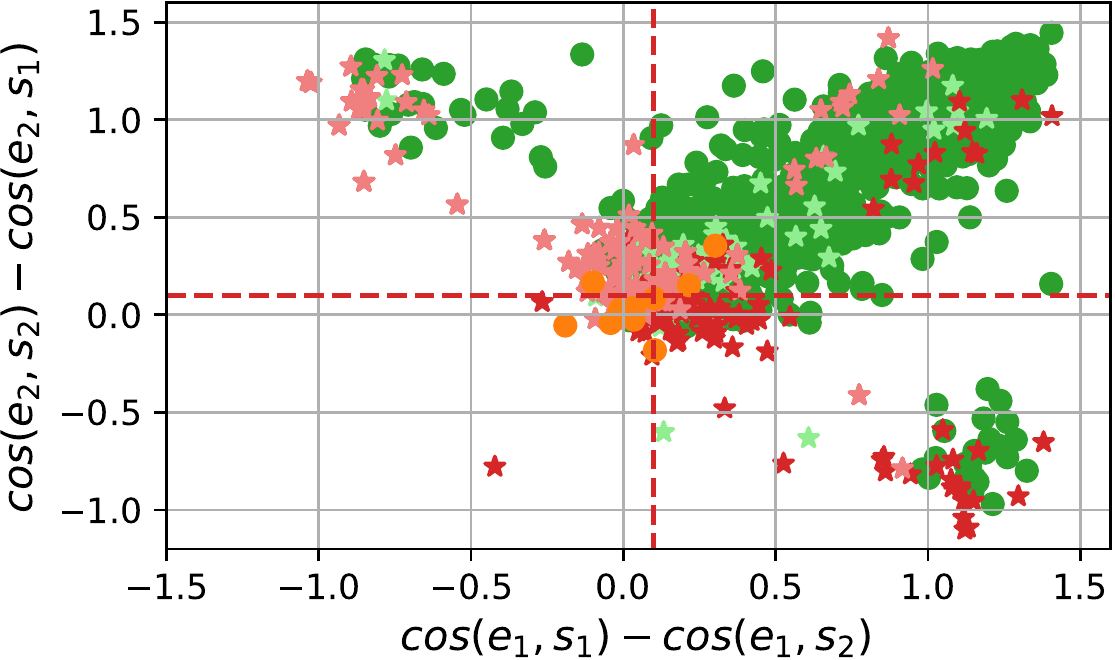}}
      \caption{Similarity analysis in embedding space. $cos(e_i, s_j)$ denotes the cosine distance between the speaker embedding of enrollment utterance $e_i$ from speaker $i$ and that of the source $s_j$ from speaker $j$. Speaker 1 is set to be the target by default in (a)(c), and speaker 2 the interferer. Dashed line in red denotes a border where two similarities are equal in (a)(c), while in (b)(d) it delimits a margin. Samples indicated by pink stars, whose SI-SDRi metrics are significant negative values, are very likely to have confused the target, which is confirmed by our listening test as well as a previous research\cite{X-TasNet}. All data are from the test set of Libri2Mix with no discard.}
      \label{fig:Analysis}
      \vspace{-1.5em} 
    \end{figure*}  
    
    In this paper, we first conduct an analysis of the \textit{target confusion problem}, emphasizing the importance of distinctive speaker embeddings for end-to-end speaker extraction. Then we explore three different metric learning methods, namely, triplet loss, prototypical loss and generalized end-to-end loss, and integrate them into the end-to-end training of a deep speaker extraction model. The key behind this is that speaker extraction is an open-set setting and we need speaker embeddings with large inter-speaker and small intra-speaker distances. Finally, to further eliminate target confusion during inference, we propose a post-filtering strategy to revise the wrong results. To be specific, we first identify confusion samples by comparing the similarities between the target source estimate and the enrollment utterance, then the true target source can be recovered by a subtraction operation. Experiments show that our methods improve the baseline by more than 1dB in terms of SI-SDRi.
    
    
\section{Target confusion problem}
\subsection{Speaker extraction}

    Given an enrollment utterance $e$, speaker extraction is to extract target speaker's voice $\hat{s}$ out of a speech mixture $y$. To make it simple, a two-speaker anechoic setup is considered in the following (i.e. $y=s_1+s_2$). Either of the speakers in the mixture can be set as the target speaker, and the other the interferer.
    
\subsection{Target confusion problem}
    
    As depicted in Figure \ref{fig:LongTail}, end-to-end speaker extraction model tends to come across with the \textit{target confusion problem} during inference, that is, the model extracts the interfering speaker instead of the target speaker, and hence generate a wrong result. This leads to a situation that end-to-end speaker extraction even performs inferior to its BSS counterpart when a similar separation network is used, despite the assistance of an additional speaker encoder and enrollment utterances. Intuitively, \textit{target confusion problem} can originate from two aspects:
    
    \noindent\textbf{utterance bias} Considering the variability of speech, an utterance may deviate from its speaker cluster where it belongs to, and even tend to an interfering speaker cluster. Such variability comes from many uncontrollable factors of the speech like emotion, intonation, prosody and even speed. We refer to this as \textit{utterance bias}. Note that this may occur in the source signal $s$ as well as the enrollment utterance $e$. 
    
    \noindent{\textbf{embedding bias}} On one hand, the network architecture of speaker encoders for end-to-end speaker extraction are generally more simple compared with those used in speaker recognition tasks\cite{ResNet-34}\cite{VGGVox}\cite{ECAPA-TDNN}, which results in limited capability in speaker characteristics modeling. On the other hand, SI-SDRi is usually set as the only loss function in the end-to-end training, which does not guarantee well distinguishable speaker embeddings; These bring about the \textit{embedding bias}, that is, the speaker embedding is not distinctive enough, such that it does not represent the target speaker accurately, or it fails to distinguish the target speaker from the interferer.
    
    
    We conduct an experiment for a further analysis and comparison. In Figure \ref{fig:Analysis}(a)(b), similarities between speakers are measured by a pretrained ECAPA-TDNN\cite{ECAPA-TDNN}, which is a state-of-the-art embedding encoder used in speaker verification with an equal error rate (EER) of less than 1\%. As shown in Figure \ref{fig:Analysis}(a), 99.7\% of the enrollment utterances are closer to their target sources instead of the interfering speech in the embedding space. And in Figure \ref{fig:Analysis}(b), for more than 98.9\% of our test cases, the two aforementioned similarities have a margin of more than 0.1. Most interestingly, only 2.4\% of target confusion samples, which is denoted with pink stars, lie beyond the border in Figure \ref{fig:Analysis}(a), and only 6.4\% of them out of the 0.1 margin in Figure \ref{fig:Analysis}(b). 
    
    Things turn out to be very different when we come to the speaker encoder in an end-to-end trained speaker extraction model. TD-SpeakerBeam was adopted for the evaluation, in which the speaker encoder is composed of an encoder layer and a convolution block\cite{TD-SpeakerBeam}. For a certain amount of samples, the enrollment utterances are much closer to the interferers instead of their target speakers, as shown in Figure \ref{fig:Analysis}(c). It is worth noting that more than 45.1\% of target confusion samples lie above the border where two similarities are equal. And in Figure \ref{fig:Analysis}(d), 65\% of target confusion samples lie out of the 0.1 margin. 
    
    Comparing above observations we can draw some conclusions. First, while \textit{utterance bias} may exist in some situations where speakers' voices are very similar, it is much less significant than expected, at least in our test case; Second, a considerable amount of target confusion samples are caused by speaker embeddings that are not distinguishable enough. We argue that \textit{embedding bias} is underestimated.
    
    

\section{Methods}
    

\subsection{Metric learning for end-to-end speaker extraction}

    In this section, we introduce how to integrate practical metric learning methods with the end-to-end training of a speaker extraction model. The essence behind this is to generate speaker embeddings with large inter-speaker and small intra-speaker distance through explicit optimization in the metric space, so that it does not confuse the target and the interferer. Three different metric learning methods are explored in the following, including triplet loss\cite{Triplet}, prototypical loss\cite{PNL}\cite{Prototypical} and generalized end-to-end loss\cite{GE2E}\cite{MetricLearning}. 
    
    \noindent\textbf{Multi-task learning} A multi-task learning framework is adopted to combine the reconstruction loss and the metric learning loss:
        \begin{equation}
            L = \beta L_{ML} + \frac{1}{N} \sum_{n=1}^N L_n
        \end{equation}
    where $N$ denotes the batch size, $L_{\textit{n}}$ and $L_{ML}$ are loss functions for waveform reconstruction and metric learning. $\beta$ is a hyperparameter. The negative scale-invariant signal-to-distortion ratio is used as reconstruction loss\cite{Conv-TasNet}.

    \noindent\textbf{Triplet loss (TL)} A triplet $(u, v, w)$ consists of an anchor $u$, a positive $v$ and a negative $w$. The triplet loss forces the encoder to reserve a margin between the distance of the anchor-positive pair $(u,v)$ and that of the anchor-negative pair $(u,w)$:
        \begin{equation}
            l_{TL}(u,v,w) = max(0, d(u,v)-d(u,w)+\alpha)
        \end{equation}
    where $d(a,b)$ denotes the $L2$ distance between $L2$-normed embeddings of utterance $a$ and $b$. $\alpha$ is the margin, which is a hyperparameter. We propose two schemes to form the triplet. In the first scheme $TL_1(s_t, e_t, e_f)$, target source $s_t$ is set as the anchor, while the enrollment utterance of target speaker and interferer are intuitively set as the positive and negative respectively; In the second scheme $TL_2(s_t, \hat{s}_t, e_f)$, target estimate $\hat{s}_t$ replaces the enrollment $e_t$ as the positive. At last, the loss is averaged over the batch: $L_{TL}=\frac{1}{N} \sum_{n=1}^N l_{TL}(u_n, v_n, w_n)$.
    
    
    
    \noindent\textbf{Prototypical loss (PL)} In prototypical loss, utterances of speaker $k$ are divided into a support set $S_k$ and a query set $Q_k$. The prototype $\textbf{r}_k$, i.e. the speaker centroid, is calculated as the mean of speaker embeddings from $S_k$:
        \begin{equation}
            \textbf{r}_k = \frac{1}{|S_k|} \sum_{x_s \in S_k} E(x_s)
        \end{equation}
    where $E()$ denotes the speaker encoder which maps an utterance $x_s$ to an embedding. The likelihood that an utterance $x_n$ in the batch belongs to its speaker $z_n$ is calculated with a softmax over all $I$ speakers in the training set:
        \begin{equation}
            p_{PL}(x_n,z_n) = \frac{e^{-d(E(x_n), \textbf{r}_{z_n})}}{\sum_i^I e^{-d(E(x_n),\textbf{r}_i)}}
        \label{eq:SoftmaxPL}
        \end{equation}
    Following the setup in $TL$, two different schemes are investigated: $PL_1(x_n=e_t)$ and $PL_2(x_n=\hat{s}_t)$. Speech for $S_k$ are from the whole training set, while those for $Q_k$ are all from the current batch. Finally, negative logarithm is applied to Eq. (\ref{eq:SoftmaxPL}) for a maximum likelihood estimation (MLE): 
        \begin{equation}
            L_{PL} = \frac{1}{|Q_k|}\sum_{x_q \in Q_k} -log(p_{PL}(x_q,z_q))
        \end{equation}
    
    \noindent\textbf{Generalized end-to-end loss (GL)} Different from $PL$, $GL$ utilizes two kinds of speaker centroids:
        \begin{equation}
            \textbf{c}_k(x_n) = \begin{cases}
                \frac{1}{|C_k|} \sum_{x_c \in C_k} E(x_c), x_n \not\in C_k \\
                \frac{1}{|C_k|-1} \sum_{x_c \in C_k, x_c \neq x_n} E(x_c), x_n \in C_k \\
            \end{cases}
        \end{equation}
    where $C_k$ is an utterance bank of speaker $k$ where speech is from the whole training set, $x_n$ is an utterance in the batch whose similarity to be measured. Similar to $PL$, likelihood is calculated with a softmax:
        \begin{equation}
            p_{GL}(x_n,z_n) = \frac{e^{w \cdot cos(E(x_n), \textbf{c}_{z_n}(x_n))+b}}{\sum_i^I e^{w \cdot cos(E(x_n),\textbf{c}_i(x_n))+b}}
        \end{equation}
    where $w$ and $b$ are learnable weights. Following the previous, we investigate two settings: $GL_1(x_n=e_t)$ and $GL_2(x_n=\hat{s}_t)$. At last, negative logarithm is applied for a MLE: 
        \begin{equation}
            L_{GL} = \frac{1}{N}\sum_n^N -log(p_{GL}(x_n,z_n))
        \end{equation}
    
\subsection{Post-filtering strategy}
\begin{figure}[t]
      \centering
      \subfloat[Samples' distribution]{
        \label{fig:PostFiltering-b}
        \includegraphics[height=2.1cm]{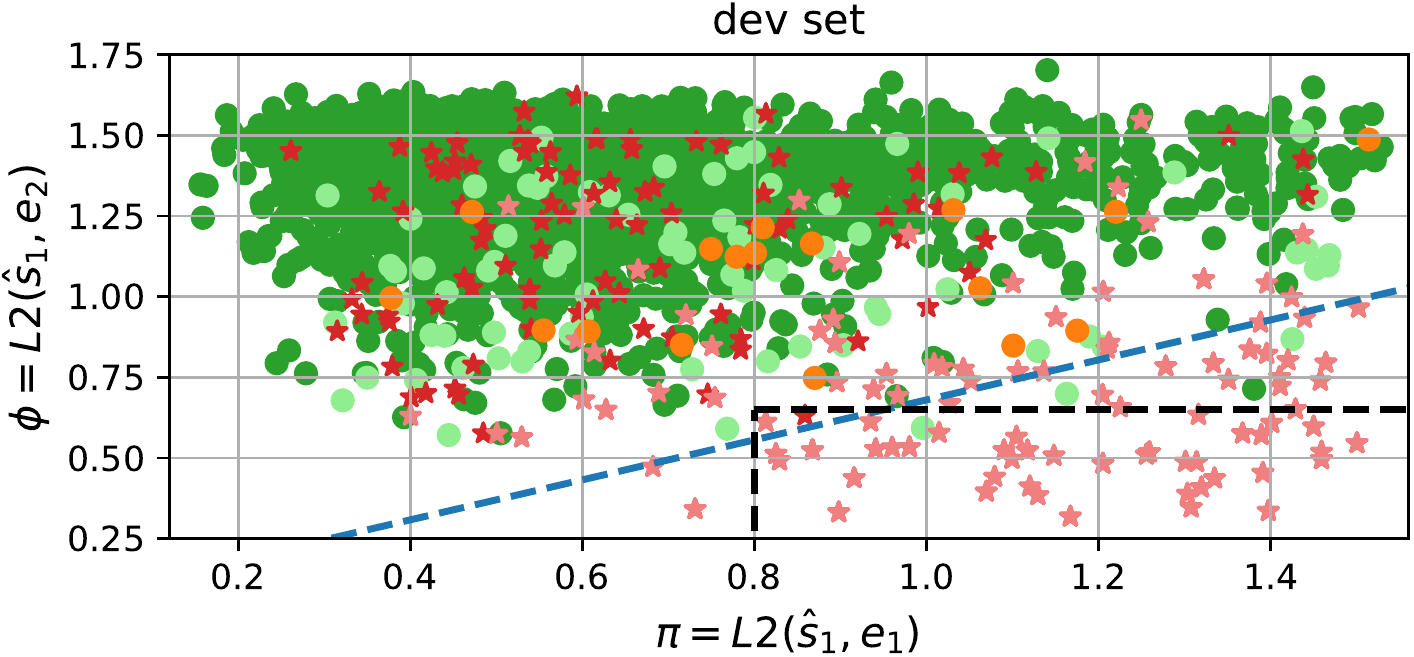}}
      \subfloat[Identified samples]{
        \label{fig:PostFiltering-c}
        \includegraphics[height=2.2cm]{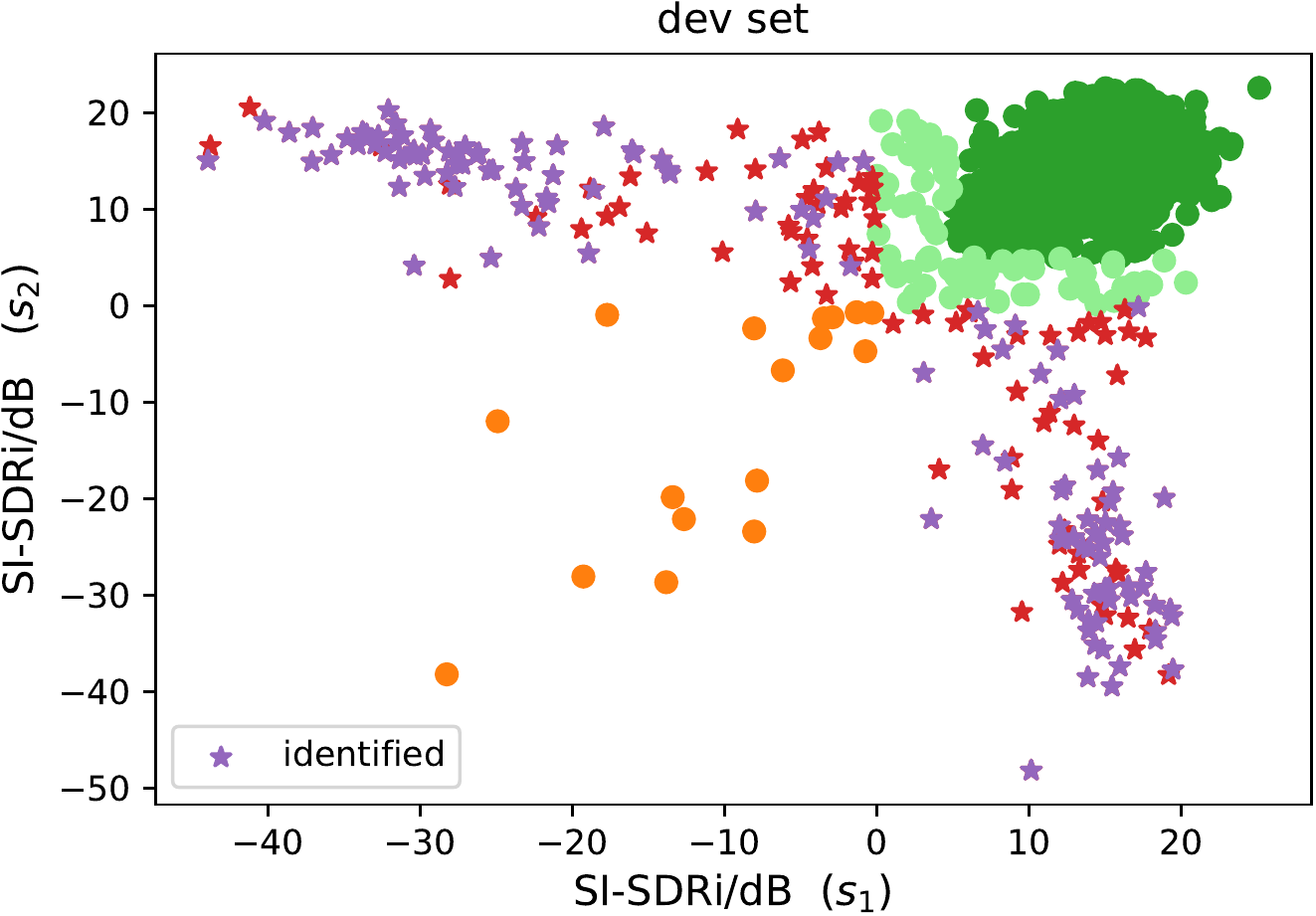}}
     \caption{Configuring the post-filtering strategy on validation set. (a) Distribution of data samples in a space spanned by $\pi=L2(\hat{s}_1,e_1)$ and $\phi=L2(\hat{s}_1,e_2)$, where $L2(a,b)$ denotes the $L2$  distance between two $L2$-normed embeddings of utterance $a$ and $b$. Rectangular and linear decision borders are depicted by black and blue dashed lines respectively. (b) Target confusion samples identified by our methods are illustrated with purple stars. All data are from the dev set of Libri2Mix with no discard.}
     \label{fig:PostFiltering}
     \vspace{-2em} 
    \end{figure}

    To further improve the robustness of the system, we propose a post-filtering strategy ($PF$) to first identify and then rectify those target confusion samples during inference. Specifically, our pipeline has three steps. The model trained with the aforementioned methods first consumes the speech mixture $y_m$ and an enrollment utterance $e_m$ to generate a target source estimate $\hat{s}_m'$; Secondly, $\hat{s}_m'$ is evaluated in two dimensions: one is its similarity with the target speaker, denoted as $\pi$, and the other is that with the interferer, denoted as $\phi$. Considering that ground-truth sources are not available during inference, speaker clusters are estimated by their enrollment speech, as illustrated in Figure \ref{fig:PostFiltering}(a). Then a decision border to classify target confusion samples can be easily obtained by solving an optimization problem in a $M$-sample discrete space spanned by $\pi$ and $\phi$. For this we propose two objective functions. One is a rectangular border $PF^{\textit{rec}}$:
				\begin{equation}
				\label{eq:rec}
				        \max_{\Pi, \Phi} \quad \sum_{m=1}^M g_m^{\textit{rec}}(\Pi, \Phi)
				    \end{equation}
				    \begin{equation}
				        g_m^{\textit{rec}}(\Pi, \Phi) = \begin{cases}
				            l(s_m, y-f(y_m|e_m)), \pi > \Pi, \phi < \Phi \\
				            l(s_m, f(y_m|e_m)), otherwise \\
				        \end{cases}
				    \end{equation}
	where $\Pi$ and $\Phi$ are threshold egparameters for $\pi$ and $\phi$ respectively. $l$ is a SI-SDRi metric\cite{SISDR}, and $f$ is the deep speaker extraction model which produces a target source estimate normalized w.r.t. the input mixture $y_m$, given an enrollment utterance $e_m$. The other is a linear border $PF^{\textit{lin}}$, such that:
				\begin{equation}
				\label{eq:linear}
				        \max_{\mu, \lambda} \quad \sum_{m=1}^M g_m^{\textit{lin}}(\mu, \lambda)
				    \end{equation}
				    \begin{equation}
				        g_m^{\textit{lin}}(\mu, \lambda) = \begin{cases}
				            l(s_m, y-f(y_m|e_m)), \phi < \mu \pi + \lambda \\
				            l(s_m, f(y_m|e_m)), otherwise \\
				        \end{cases}
				    \end{equation}
    where $\mu$ and $\lambda$ are parameters to be tuned. There are many training-free ways to solve Eq. (\ref{eq:rec})(\ref{eq:linear}), e.g. brute-force searching. Note that the test set is assumed to be inaccessible for the tuning, and parameters are configured only using the validation set where ground-truth sources are available. Lastly, target confusion samples can be identified during inference, by thresholding on $\pi$ and $\phi$ with tuned parameters $\Pi$ and $\Phi$ (or with $\mu$ and $\lambda$), as depicted in Figure \ref{fig:PostFiltering}(b). Under the two-speaker anechoic setup introduced in Section 2.1, those identified samples are inverted by being subtracted from the mixture $y$, after which the true target source is recovered:
        \begin{equation}
            \hat{s}_m = y_m-\hat{s}_m'
        \end{equation}
    where $\hat{s}_m$ is the final output after the post-filtering. More complex scenarios like multiple speakers ($\#spk\ge3$) or noisy environment will be explored in the future.
    
\section{Experiments}

\begin{figure*}[t]
      \centering
      \subfloat[$NS$ baseline]{
        \label{fig:result-1}
        \includegraphics[height=2.7cm]{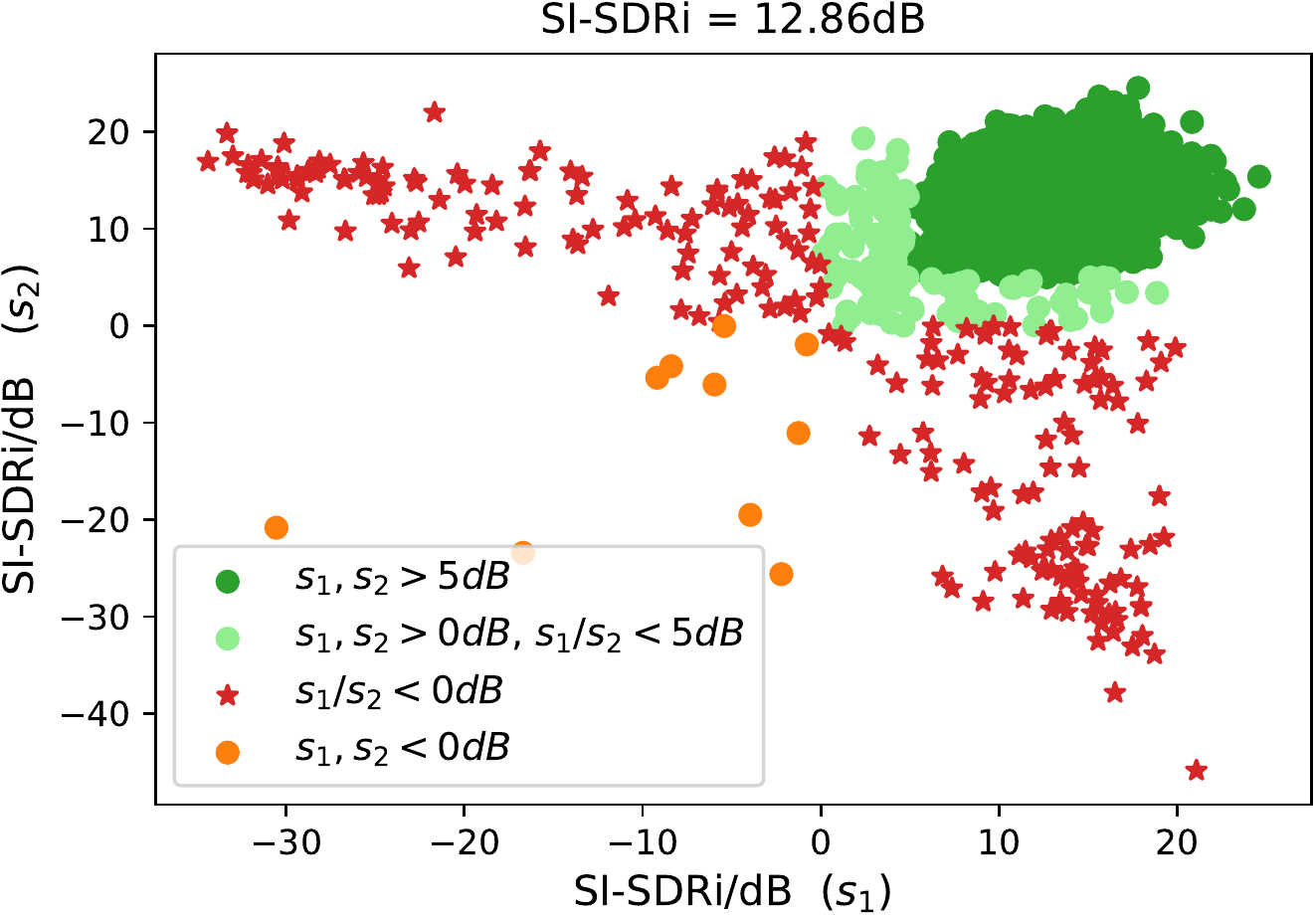}}
      \subfloat[Proposed $PL_2$]{
        \label{fig:result-2}
        \includegraphics[height=2.7cm]{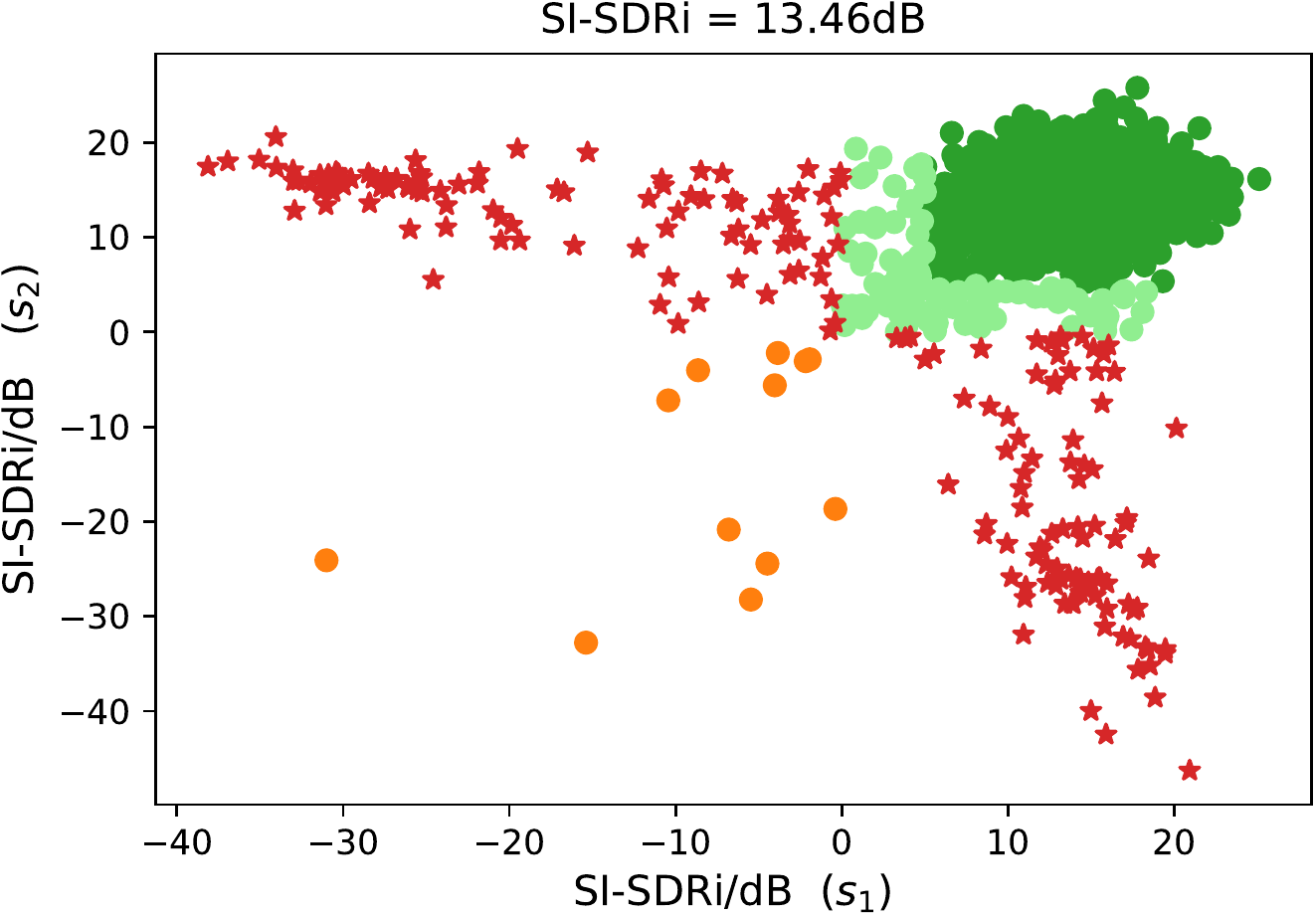}}
      \subfloat[Proposed $PL_2+PF^{\textit{lin}}$]{
        \label{fig:result-3}
        \includegraphics[height=2.7cm]{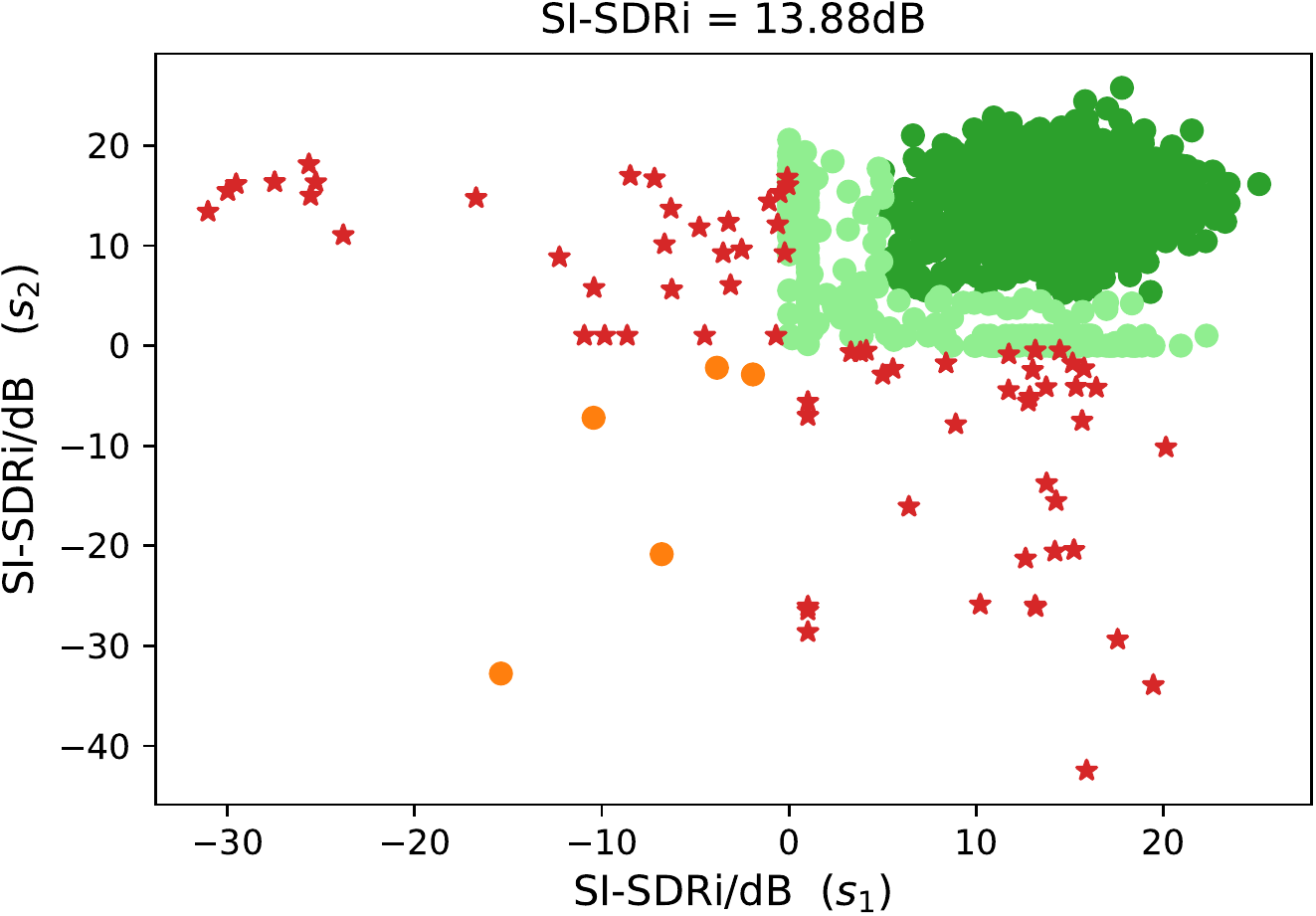}}
      \subfloat[statistic]{
        \label{fig:result-4}
        \includegraphics[height=2.7cm]{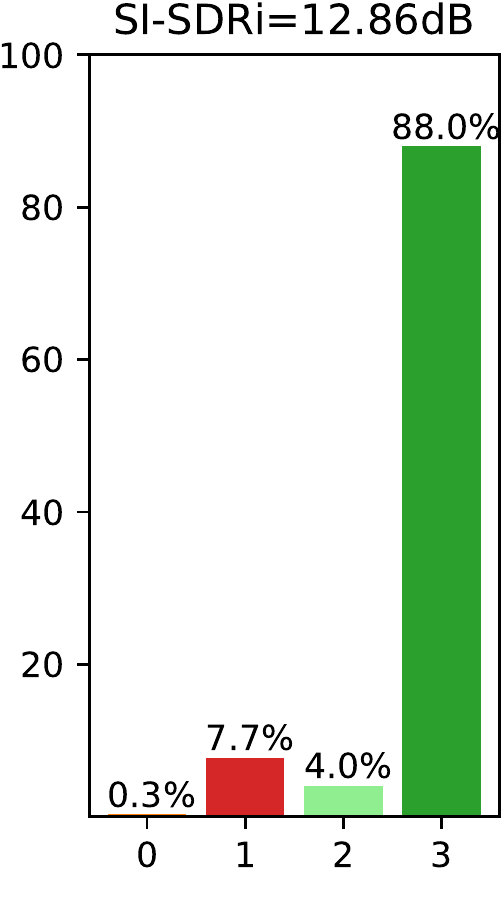}}
      \subfloat[statistic]{
        \label{fig:result-5}
        \includegraphics[height=2.7cm]{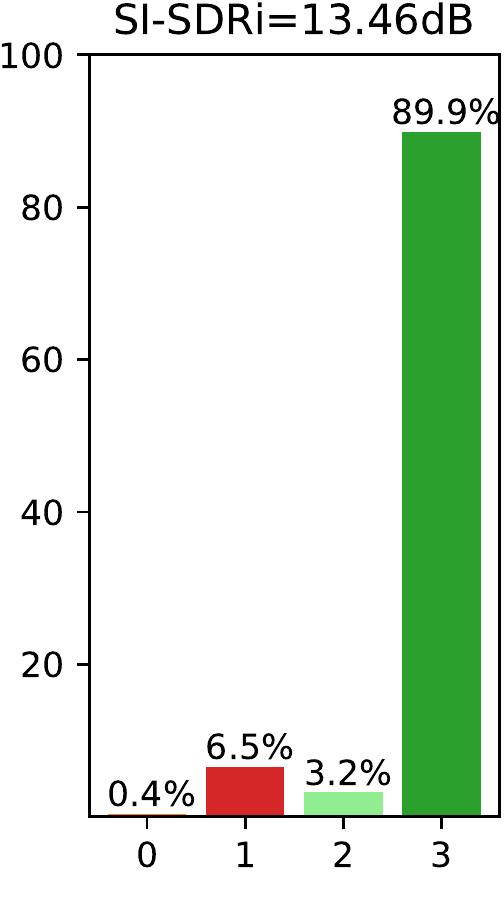}}
      \subfloat[statistic]{
        \label{fig:result-6}
        \includegraphics[height=2.7cm]{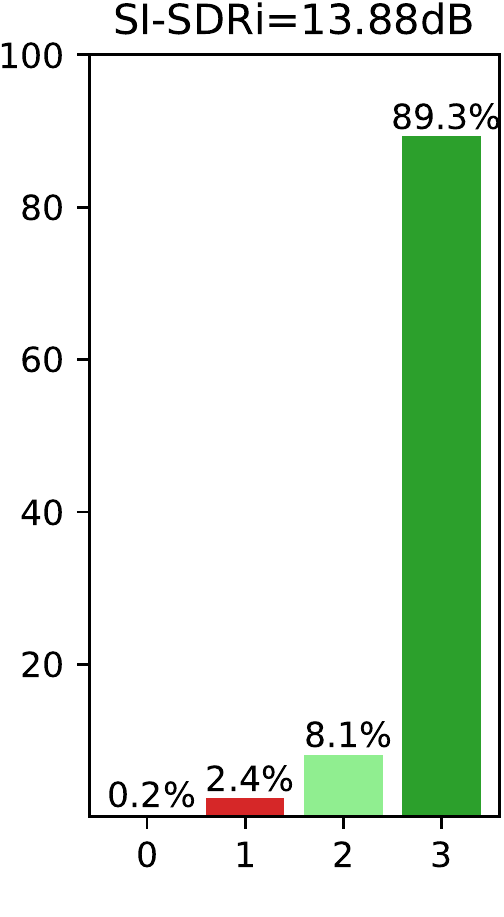}}
     \caption{Joint distributions of two speakers' SI-SDRi performance on the test set of Libri2Mix. In each mixture audio, two speakers are set to be the target in turns. (a)(b)(c) are the joint distributions, and (d)(e)(f) are their corresponding statistical histograms. } 
     \vspace{-2em} 
     \label{fig:JointDistribution}
    \end{figure*}
    
\subsection{Preparation}
    The TD-SpeakerBeam\cite{TD-SpeakerBeam} is adopted for our experiments. It is chosen such that we can fairly compare it with its BSS counterpart Conv-TasNet\cite{Conv-TasNet}. These two models share the same separation network, except for that TD-SpeakerBeam has an additional speaker encoder and an embedded adaptation layer. We validate our methods on the popular LibriMix\cite{LibriMix} dataset. The \textit{train-100} subset is used for training, \textit{dev} subset for configuring the post-filtering as well as for the validation set during training, while \textit{test} is used for the final evaluation. All speech audios are 8kHz and mixtures are in 'minimum' mode. During training, both input mixtures and enrollment speech are randomly truncated to 3 seconds, while full-length audios are used for testing.
    
\subsection{Results}
    We compare the proposed training methods with three baselines on the \textit{sep\_clean} task of Libri2Mix: (1) $NS$: negative SI-SDR as the only training target; (2) $CE$: multi-task learning with a mulit-class cross-entropy loss for speaker classification and a negative SI-SDR loss for waveform approximation; (3) $BSS$: Conv-TasNet trained with permutation-invariant training\cite{PIT}\cite{uPIT}.  
    
\begin{table}[h]
\centering
\begin{tabular}{@{}cccc@{}}
\toprule
\textbf{}                             & \textbf{SI-SDRi(dB)} & \textbf{PESQ}                      & \textbf{params}          \\ \midrule
\multicolumn{1}{c|}{$NS$}             & 12.86                & \multicolumn{1}{c|}{2.75}          & -                        \\
\multicolumn{1}{c|}{$CE$}             & 13.05                & \multicolumn{1}{c|}{2.78}          & $\beta$=0.2              \\
\multicolumn{1}{c|}{$BSS$}            & 13.40                & \multicolumn{1}{c|}{2.74}          & -                        \\ \midrule
\multicolumn{1}{c|}{$TL_1$}           & 13.31                & \multicolumn{1}{c|}{2.82}          & $\beta$=0.2, $\alpha$=1  \\
\multicolumn{1}{c|}{$TL_2$}           & 13.36                & \multicolumn{1}{c|}{2.83}          & $\beta$=0.2, $\alpha$=1  \\ \cmidrule(l){2-4} 
\multicolumn{1}{c|}{$PL_1$}           & 13.46                & \multicolumn{1}{c|}{2.85}          & $\beta$=0.2, $|S_k|$=5   \\
\multicolumn{1}{c|}{$PL_2$}           & 13.46                & \multicolumn{1}{c|}{2.85}          & $\beta$=0.1, $|S_k|$=5   \\ \cmidrule(l){2-4} 
\multicolumn{1}{c|}{$GL_1$}           & \textbf{13.47}       & \multicolumn{1}{c|}{\textbf{2.85}} & $\beta$=0.1              \\
\multicolumn{1}{c|}{$GL_2$}           & 13.44                & \multicolumn{1}{c|}{2.83}          & $\beta$=0.1              \\ \midrule
\multicolumn{1}{c|}{$NS+PF^{\textit{rec}}$}    & 13.13                & \multicolumn{1}{c|}{2.76}          & $\Pi$=0.4, $\Phi$=0.4    \\
\multicolumn{1}{c|}{$NS+PF^{\textit{lin}}$}    & 13.14                & \multicolumn{1}{c|}{2.76}          & $\mu$=0.4, $\lambda$=0.2 \\
\multicolumn{1}{c|}{$CE+PF^{\textit{rec}}$}    & 13.31                & \multicolumn{1}{c|}{2.79}          & $\Pi$=0.5, $\Phi$=0.5    \\
\multicolumn{1}{c|}{$CE+PF^{\textit{lin}}$}    & 13.32                & \multicolumn{1}{c|}{2.79}          & $\mu$=0.4, $\lambda$=0.2 \\
\multicolumn{1}{c|}{$PL_2+PF^{\textit{rec}}$}  & 13.82                & \multicolumn{1}{c|}{2.85}          & $\Pi$=0.8, $\Phi$=1.0    \\
\multicolumn{1}{c|}{$PL_2+PF^{\textit{lin}}$}  & \textbf{13.88}       & \multicolumn{1}{c|}{\textbf{2.86}} & $\mu$=0.6, $\lambda$=0.3 \\ \bottomrule
\end{tabular}
\caption{Comparing the overall performance.}
\label{tab:exp}
\vspace{-2em}
\end{table}

    Results are presented in terms of SI-SDRi\cite{SISDR} and PESQ\cite{PESQ} in Table \ref{tab:exp}. For the sake of space, only the best results are reported, together with their hyperparameters. As illustrated in the first and the third row, the deep speaker extraction model is inferior to its BSS counterpart by 0.54dB in terms of SI-SDRi, which is consistent with our statements in Section 1. By observing row four to row nine, we can see that all metric learning methods promote the performance and outperform the $CE$ baseline. The performance difference between scheme 1 and  scheme 2 is minor. Among proposed training methods, $TL$ performs the worst (13.36dB SI-SDRi). $PL$ and $GL$ achieve similar results, improving the performance by 0.6dB and 0.61dB SI-SDRi respectively, and both of them outperform the $BSS$ baseline. 
    
\begin{figure}[h]
      \centering
      \includegraphics[width=\linewidth]{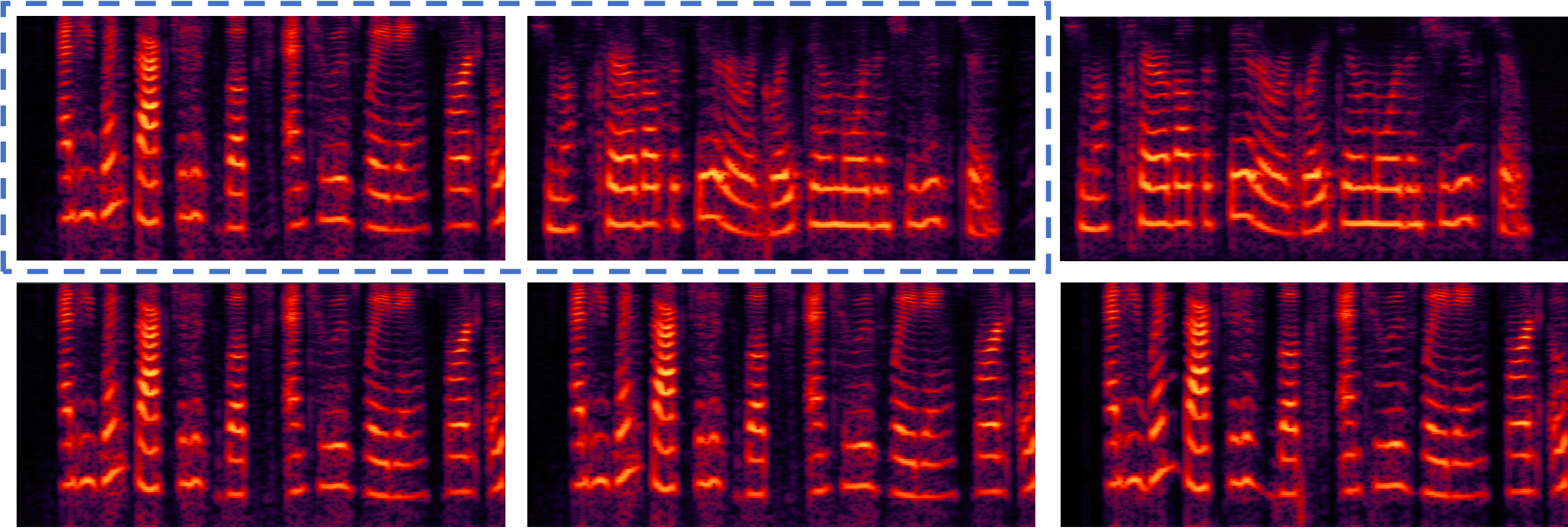}
      \caption{Spectrograms from a data sample. In the first column are the original estimates from the deep model, in the second column are final outputs after post-filtering, and the ground truths are in the last column. As shown in the blue dashed box, target confusion is rectified by our post-filtering strategy.}
      \label{fig:rectify}
    \vspace{-1em} 
    \end{figure}

    For the post-filtering strategy, threshold parameters are tuned on \textit{dev} set in advance and set to be constant during inference. Note that the threshold parameters should not be too precise (e.g. one decimal place would be fair enough) to avoid overfitting on the validation set. As shown in the last six rows in Table \ref{tab:exp}, both $PF^{\textit{rec}}$ and $PF^{\textit{lin}}$ further improve the performance of baseline $NS$ and $CE$, as well as our $PL_2$. An example\footnote{More audio examples are available at our demo webpage:  https://zhazhafon.github.io/demo-confusion/} is depicted in Figure \ref{fig:rectify}. Interestingly, applying $PF$ on the basis of proposed training methods brings more gain in SI-SDRi than simply applying it to the baselines, and it further advances our results by 0.36dB and 0.42dB respectively. This is due to that proposed training methods provide more reliable speaker embeddings and thus form a more distinctive decision border in the subspace spanned by $\pi$ and $\phi$, which is vital for the post-filtering.
    
     We visualize some of the results in Figure \ref{fig:JointDistribution}. The proposed methods significantly alleviate the long-tail distribution in end-to-end speaker extraction. The best performance is achieved by $PL_2+PF^{\textit{lin}}$, with a SI-SDRi of 13.88dB and a PESQ of $2.86$.
    
\section{Conclusions}
    
    In this paper, we conduct an analysis of what we refer to as the \textit{target confusion problem} in end-to-end speaker extraction, and proposed to solve it with metric learning methods and a post-filtering strategy. Experiments show that our methods promote the performance by more than 1dB SI-SDRi. In future work, we plan to extend our methods to more complicated scenarios, for example, multi-talker ($\#spk\ge3$) and noisy extraction.
    



\input{template.bbl}

\bibliographystyle{IEEEtran}


\end{document}